\input phyzzx.tex
\tolerance=1000
\voffset=-0.3cm
\hoffset=1.0cm
\def\rl{\rightline}

\def\t1{{\tilde 1}}

\def\t m^2{\tilde m^2}

\REF{\MAL} {J. M. Maldacena, hep-th/9711200.}
\REF{\KLEB} {S. S. Gubser, I. R. Klebanov and A. W. Peet.}
\REF{\kleb2} {I. R. Klebanov, hep-th/9702076}
\REF{\kleb3} {S. S. Gubser, I. R. Klebanov and A. A. Tseytlin,
hep-th/9703040.}
\REF{\kleb4} {S. S. Gubser and I. R. Klebanov,  hep-th/9708005.}
\REF{\MS} {J. M. Maldacena and A. Strominger, hep-th/9710014.}
\REF{\POL} {A. M. Polyakov,  hep-th/9711002.}
\REF{\GKP} {S.S. Gubser, I.R. Klebanov and A.M. Polyakov,
hep-th/9802109.}
\REF{\WIT} {E. Witten, hep-th/9802150.}
\REF{\0}  {K. Sfetsos and K. Skenderis, hep-th/9711138.}
\REF{\KKR} {R. Kallosh, J. Kumar and A. Rajaraman, hep-th/9712073}
\REF{\CKKTV} {P. Claus, R. Kallosh, J. Kumar, P. Townsend and A. Van
Proeyen,
hep-th/9801206.}
\REF{\FF}  {S. Ferrara and C. Fronsdal,  hep-th/9712239.}
\REF{\2} {N. Itzhaki, J. M. Maldacena, J. Sonnenschein and S.
Yankielowicz,
hep-th/9802042.}
\REF{\3} {M. Gunaydin and D. Minic,  hep-th/9802047.}
\REF{\4} {G. T. Horowitz and H. Ooguri, hep-th/9802116.}
\REF{\KS} {S. Kachru and E. Silverstein, hep-th/9802183.}
\REF{\6} {M. Berkooz,  hep-th/9802195.}
\REF{\65} {V. Balasubramanian and F. Larsen,  hep-th/9802198.}
\REF{\7}  {S. J. Rey and J. Yee, hep-th/9803001.}
\REF{\8} {J. M. Maldacena, hep-th/9803002.}
\REF{\9} {M. Flato and C. Fronsdal, hep-th/9803013.}
\REF{\LNV}  {A. Lawrence, N. Nekrasov and C. Vafa, hep-th/9803015.}
\REF{\newKleb} {S. S. Gubser, A. Hashimoto, I. R. Klebanov and M.
Krasnitz,
hep-th/9803023.}
\REF{\AV} {I. Ya. Aref'eva and I. V. Volovich, hep-th/9803028.}
\REF{\CCD}{ L. Castellani, A. Ceresole, R. D'Auria, S. Ferrara, P. Fre'
and M. Trgiante, hep-th/9803039.}
\REF{\FFZ} {S. Ferrara, C. Fronsdal and A. Zaffaroni,  hep-th/9802203.}
\REF{\YAR}{O. Aharony, Y. Oz and Z. Yin, hep-th/9803051.}
\REF{\KRN} {H. J. Kim, L. J. Romans and P. van Nieuwenhuizen, Phys. Rev.
D32 (1985)
389.}
\REF{\BCERS} {B. Biran, A. Casher, F. Englert, M. Rooman and P. Spindel,
Phys. Lett. {\bf 134B} (1984) 179.}
\REF{\GUN} {M. Gunaydin and N. P. Warner,  Nucl. Phys. B272 (1986)
99-124.}
\REF{\CENR} {A. Casher, F. Englert, H. Nicolai and M. Rooman,  Nucl. Phys.
B243 (1984)
173-188.}
\REF{\VAN} {P. Van Nieuwenhuizen, Class. Quantum Grav. {\bf 2} (1985) 1.}
\REF{\MIN}{S Minwalla, hep-th/9803053.}
\REF{\VANN} {L. Castellani, R. D'Auria, P. Fre, K. Pilch and
 P. Van Nieuwenhuizen,Class. Quantum Grav. {\bf 1} (1984) 339.}
\REF{\GUNA} {M. Gunaydin, P. van Nieuwenhuizen and N. P. Warner,  Nucl.
Phys. B255 (1985) 63-92.}
\REF{\PNT} {K. Pilch, P. van Nieuwenhuizen and P. K. Townsend,  Nucl.
Phys. B242 (1984)
377-392.}

\singlespace
\rl{SU-ITP-98/12}
\rl{\today}
\rl{hep-th/9803077}
\pagenumber=0
\normalspace
\medskip
\bigskip
\titlestyle{\bf{ Supergravity on $AdS_{4/7} \times S^{7/4}$ and M Branes}}
\smallskip
\author{ Edi Halyo{\footnote*{e--mail address:
halyo@dormouse.stanford.edu}}}
\smallskip
\centerline {Department of Physics}
\centerline{Stanford University}
\centerline {Stanford, CA 94305}
\smallskip
\vskip 2 cm
\titlestyle{\bf ABSTRACT}

We calculate the dimensions of  operators in three and six dimensional
superconformal field theories by using the duality between these theories
at large $N$ and $D=11$ supergravity on $AdS_{4/7} \times S^{7/4}$. We
find that for the duality relations to work
the Kaluza--Klein masses given in the supergravity literature must be
rescaled and/or shifted.

\singlespace
\vskip 0.5cm
\endpage
\normalspace

\centerline{\bf 1. Introduction}

Recently a duality between some superconformal brane world--volume
theories and
supergravity on corresponding AdS spaces was conjectured in Ref. [\MAL].
More
precisely it was argued that for  D3 branes in a certain limit near the
branes the large $N$ world--volume dynamics is dual to IIB supergravity on
$AdS_5 \times S^5$. This background is precisely the near horizon geometry
of the D3 branes and one can probe it by supergravity because
the radius of the $AdS$ space and the sphere is large (proportional to a
positive power of $N$ times the Planck length) 
or the curvature is small. The geometry near the D3 branes is also given
by the $S^5$
compactification of IIB supergravity which is maximally supersymmetric,
i.e. gauged supergravity.
Using this duality the isometry of the $AdS_5$ space
$SO(3,2)$ becomes the conformal symmetry on the brane theory whereas the
isometry of $S^5$ which is $SO(6)$ becomes the R symmetry. Results about
entropy of and
scattering from black holes made of D3 branes seem to
support this conjecture [\KLEB-\MS].
The conjecture can also be applied
to other non--dilatonic branes, i.e. the membranes and fivebranes of
$D=11$ supergravity
(or M theory).
In that case the duality is between large $N$ world--volume theories of
M2/M5 branes and $D=11$
supergravity (M theory) on $AdS_{4/7} \times S^{7/4}$.
Recently other related work appeared in [\POL-\CCD].

A much more precise formulation of this duality was given in Ref. [\WIT].
There
it was shown that supergravity on an $AdS_{d+1}$ background is equivalent
to a
superconformal field theory (SCFT) on the boundary of the $AdS_{d+1}$
space
which is
$d$ dimensional Minkowski space $M_d$. To every supergravity field on
$AdS_{d+1}$ there is a corresponding
chiral operator in the  SCFT on the boundary. For scalars, if the field is
massive, massless or
tachyonic the operator in the SCFT is irrelevant, marginal and relevant
respectively. Moreover, the dimension of the operator is fixed by the mass
of the field in supergravity. As a result this duality can be checked by
comparing the spectrum of supergravity on $AdS_{d+1}$ backgrounds and
dimensions of the operators
in the corresponding SCFT living on $M_d$.
The spectrum of IIB supergravity on $AdS_5 \times S^5$ is known[\KRN] and
matches
precisely the dimensions of operators of ${\cal N}=4$ $D=4$ super
Yang--Mills
theory which is the world--volume theory of D3 branes[\WIT].

In this letter we repeat this for the dualities between large $N$
world--volume theories
of M2/M5 branes and supergravity on $AdS_{4/7} \times
S^{7/4}$.{\footnote*{While this work 
was being completed [\YAR] which has the same results appeared.}} Section
2 is
a short review of the results of Ref. [\WIT] that we need and their
relation to
previous results in the supergravity literature.
In sections 3 and 4 we calculate the dimensions of the relevant and
marginal
operators in $D=3$ and $D=6$ SCFT with sixteen supercharges
using the spectrum of the $D=11$ supergravity compactified on $S^7$ and
$S^4$
respectively. Section 5
is a short discussion of our results.

\bigskip
\centerline{\bf 2. Supergravity on $AdS_{d+1}$}

In Ref. [\WIT] it was shown that supergravity on $AdS_{d+1}$ is equivalent
to
a SCFT on the boundary $M_d$. The
isometry group $SO(d,2)$ of the $AdS_{d+1}$ space is also the conformal
symmetry
of the $d$ dimensional boundary SCFT. For
a scalar field $\phi$ which has a value $\phi_0$ on the boundary there is
a
coupling of the form $\int \phi_0 {\cal O}$ to an operator $\cal O$ on the
boundary $M_d$.
The conformal dimension of $\cal O$ is related to the mass of $\phi$ by 
$${\t m^2}=\Delta(\Delta-d) \eqno(1)$$
This shows that scalar fields which are massive, massless and tachyonic
correspond to irrelevant, marginal and relevant operators on the boundary.
For a $p$ form field in supergravity the relation becomes
$${\t m^2}=(\Delta+p)(\Delta+p-d) \eqno(2)$$

Since we will use these formulas and the spectrum of $D=11$ supergravity
on
$AdS_{4/7} \times S^{7/4}$ to obtain the dimensions of operators $\cal O$
it is
important to review also the supergravity conventions. Supergravity on
$AdS_{d+1}$ has the isometry group $SO(d,2)$ which is the conformal group
of the boundary theory. This has a maximal subgroup of $SO(d) \times
SO(2)$.
The eigenvalues of the $SO(2)$ factor are denoted by $E_0$ and give the
dimension of operators in the SCFT on the boundary[\FF]. These operators
are 
bilinear currents of the boundary theory which correspond to the
supergravity fields in the bulk.
For example, in the $AdS_5$
case one has for scalars 
$$ m^2=E_0(E_0-4) \eqno(3)$$
which is exactly eq. (1). Therefore in this case $E_0=\Delta$ and $\t
m^2=m^2$.
(Throughout the paper
${\t m^2}$ refers to the mass which appears in the definition of $\Delta$
whereas $m^2$
refers to the mass which appears in the supergravity literature and enters
in the definition fo $E_0$.)
We will see that this is not always the case for $E_0$ which appear in
the supergravity literature, i.e. in some cases $E_0 \not= \Delta$ if one
identifies
${\t m^2}$ with $m^2$. However,
both $E_0$ and $\Delta$ give the conformal dimension of an operator $\cal
O$ so they
must be equal. Requiring this equality gives a relation between $m^2$ and
$\t m^2$, it turns out that the spectrum of spherically compactified
supergravity which
appears in the literature must be rescaled and/or shifted in these cases
to make contact with Ref. [\WIT]. In addition, the formula for $p$ forms
must
be modified. This is because in supergravity the mass is related to a
Maxwell operator whereas
in [\WIT] it is related to a Laplacian which is different.
For example, in the $AdS_5$ case for one forms the supergravity
result is
$$m^2=(E_0-1)(E_0-3) \eqno(4)$$
Therefore eq. (2) must be changed to
$$\t m^2=(\Delta-p)(\Delta+p-d) \eqno(5)$$

\bigskip
\centerline{\bf 3. $D=11$ Supergravity on $AdS_4 \times S^7$ and M2
branes}

In this section we consider $D=11$ supergravity on $AdS_4 \times S^7$ and
its
dual large $N$ world--volume theory of M2 branes which is a $D=3$ ${\cal
N}=8$
SCFT with sixteen supercharges. This is the world--volume theory of $N$ M2
branes
or equivalently the world--volume theory of $N$ D2 branes at infinitely
strong coupling.
In contrast to those of $N$ D2 branes (at finite coupling)
the degrees of freedom of $N$ M2 branes are not known.
The isometry group of $AdS_4$ is
$SO(3,2)$ which is also the conformal group of the M2 brane theory. The
isometry group of $S^7$ which is $SO(8)$ becomes the R symmetry of the
boundary theory.

The spectrum of $D=11$ supergravity on $AdS_4 \times S^7$ is given in Ref.
[\BCERS,\GUN,\CENR].
There are nine towers of KK states: five scalars, three vectors and a
symmetric
tensor which includes the
graviton. The five dimensional supergravity multiplet is given by a
scalar, a pseudoscalar and a vector in addition to the graviton. These
have
supergravity masses
$$m^2_{0^+}=(k-3)^2-1 \qquad m^2_{0^-}=k^2-1 \qquad m^2_1=k^2-1 \eqno(6)$$
where $k\geq 2$ for the scalar and $k \geq 1$ for the other two cases.
These are the only KK towers which include massless and tachyonic fields
which correspond
to the only marginal and relevant operators of the three dimensional SCFT
on the boundary.

However, as mentioned above before these expressions for the masses can be
used
one has to impose $E_0=\Delta$. In this case for all bosons[\CENR]
$$E_0={3 \over 2}+ {1 \over 2} \sqrt{1+m^2} \eqno(7)$$
This has to be equal to
$$\Delta={3 \over 2}+{1 \over 2} \sqrt{9+\t m^2} \eqno(8)$$
We find that the supergravity masses for the scalars must be transformed
to
$$\t m^2_{0^+}={1 \over 4}(m^2-8) \eqno(9)$$
so that the Eq.  (1) for $\Delta$ can be applied. Note that the masses are
both rescaled and shifted in this case.
The scalar masses become
$$\t m^2_{0^+}={1 \over 4}(k-4)(k-2)-2 \qquad k\geq2 \eqno(10)$$
This corresponds to operators with dimension $\Delta=k/2$. We find that
there are relevant operators of dimension $1,3/2,2,5/2$ and a marginal
operator of dimension $3$.
These operators transform
as the 35 representation of the R symmetry group $SO(8)$ which is the
symmetric traceless
representation.
Similarly the pseudoscalar masses become
$$\t m^2_{0^-}={1 \over 4}(k^2-1)-2 \qquad k\geq 1 \eqno(11)$$
This corresponds to operators with $\Delta=3/2+k/2$. In this case there
are two relevant operators of dimension $2,5/2$ and one marginal operator.
These are also in the 35 of $SO(8)$.
For the vectors using eq. (5) with $p=1$ we find
$$\Delta={3 \over 2}+ {1 \over 2} \sqrt{1+{\t m^2}} \eqno(12)$$
Comparing this with $E_0$ we get $\t m^2={m^2 / 4} $ so that
$${\t m^2}_1={1 \over 4} (k^2-1)-2 \qquad k \geq 1 \eqno(13)$$
Note that in this case there
is no shift. This leads to the dimensions $\Delta=3/2+k/2$ for the
operators
which couple to vectors. In this case the operators are in the 28 of
$SO(8)$ which is the
antisymmetric representation.

Above we found the dimensions and the R charges of the marginal and
relevant operators
in the $D=3$ SCFT with sixteen supercharges at alrge$N$ using the duality
conjectured in [\MAL]
and the relations given in [\WIT]. This was done without the knowledge of
the
fundamental degrees of freedom of the world--volume theory of $N$ M2
branes. (However
in [\MIN] it was argued that these operators can be built from scalars and
fermions living on a      
the world--volume of  $N$ M2 branes.)

\bigskip
\centerline{\bf 4. $D=11$ Supergravity on $AdS_7 \times S^4$ and M5
branes}

In this section we consider $D=11$ supergravity on $AdS_7 \times S^4$. The
dual
theory is the large $N$ limit of M5 brane world--volume theory which is
the non--Abelian tensor theory with $(0,2)$ supersymmetry. This can also
be seen as the 
world--volume theory of $N$ NS5 branes at infinite coupling. In this case
we  know
the degrees of freedom of  niether $N$ NS5  nor M5 brane world--volume
theories.
The isometry groups of
$AdS_7$ and $S^4$ which are $SO(6,2)$ and $SO(5)$ are the conformal and R
symmetries of the $D=6$ world--volume theory.

The spectrum of $D=11$ supergravity on $AdS_7 \times S^4$ is given in Ref.
[\VANN,\GUNA
,\PNT].
There are seven towers of bosonic KK states: two scalars, two vectors,
two two forms and a symmetric tensor tower which includes the graviton.
Among
these
only two, one scalar and one vector KK tower can have massless and/or
tachyonic
states. Consider first the scalars with supergravity masses
$$m^2=k(k-3) \qquad k\geq 2 \eqno(14)$$
$E_0$ in this case is given by[\GUNA]
$$E_0=3+{1 \over 2} \sqrt{36+16m^2} \eqno(15)$$
whereas from eq. (1) we get
$$\Delta=3+{1 \over 2} \sqrt{36+4 {\t m^2}} \eqno(16)$$
We find that we need to rescale the supergravity masses, ${\t m^2}=4m^2$
so that now
$${\t m^2}_0=4k(k-3) \qquad k\geq 2 \eqno(17)$$
Using this we find for the dimension of the operators $\Delta=2k$. Thus
there is one relevant
operator of dimension $4$ and one marginal operator coming from the
scalars.
For the vectors the supergravity masses are
$$m^2=k^2-1 \qquad k\geq 1 \eqno(18)$$
$E_0$ is given by (this includes an extra shift which is required to have
a well-defined number operator)
$$E_0=3+{1 \over 2} \sqrt{16+16m^2} \eqno(19)$$
Using eq. (5) for $\Delta$ for a one form we find that again ${\t
m^2}=4m^2$ so
that
$${\t m^2}_1=4(k^2-1) \qquad k \geq 1 \eqno(20)$$
which gives $\Delta=3 +2k $. We find that the vectors give one relevant
operator of
dimension $5$.
We found the dimensions of the marginal and relevant operators of the
$D=6$
SCFT  with sixteen supercharges at large $N$ using the duality of [\MAL]
and [\WIT]. As in the $D=3$ case we do not know the fundamental degrees of
freedom on the brane world--volume 
of $N$ M5 branes (however see [\MIN] again
for a possible description of these).

\bigskip
\centerline{\bf 5. Conclusions}

In this letter we calculated the dimensions of marginal and relevant
operators in three and six
dimensional SCFT with sixteen supercharges. This was done by using the KK
spectrum of
$D=11$ supergravity compactified on $S^4$ and $S^7$ and using the results
of [\WIT].
We found that if  the conformal weight of operators $\Delta$ in [\WIT] is
identified with the dimension of the boundary operators $E_0$ in the
supergravity literature one finds that  the KK masses $m^2$ and $\t m^2$
in the two cases are not  equal. In particular the masses in the
supergravity literature must be rescaled and/or shifted in order to be
used in relations given
in [\WIT].  The rescaling of the masses can be explained as
follows.\footnote*{This was pointed out to me by A. Brandhuber and N.
Itzhaki and also by R. Kallosh.} 
In ref. [\WIT] the radius
of the $AdS_{d+1}$ spacetime and the sphere were taken to be the same for
all $d$. On the
other hand, in supergravity this is only true for $AdS_5$; for the other
cases we investigated
in this paper one has $R_{AdS_{7}}=2R_{S^4}$ and $R_{AdS_{4}}=R_{S^7}/2$.
As a result,
the supergravity masses must be rescaled (by factors of $4$ and $1/4$)
before formulas of ref. [\WIT] can be used. The shift
in the scalar masses for the $AdS_4$ case is due to the conformal mass
factor for the scalars in four dimensional gravity which gives $m^2 \to
m^2+R/6$.

\bigskip
\centerline{\bf Acknowledgements}
We would like to thank Renata Kallosh, Jason Kumar and Arvind Rajaraman
for very useful discussions.

\vfill

\refout
\bye

\end